\documentclass[12pt]{article}
\usepackage{latexsym,amssymb,amsmath,amsfonts,pictex}
\newcommand{\nt}{\not \negthinspace}
\newcommand{\Nt}{\not \negthickspace}
\newcommand{\ol}[1]{{\overline{#1}}}

\begin{document}

\title{Symmetry of massive Rarita-Schwinger fields}

\author{Terry Pilling\footnote{terry@member.ams.org}\\
Department of Physics,\\ 
North Dakota State University,\\
Fargo, ND 58105-5566}

\date{\today}

\maketitle

\begin{abstract}

We derive the general lagrangian and propagator for a vector-spinor 
field in $d$-dimensions and show that the physical observables are
invariant under the so-called point transformation symmetry. 
Until now the symmetry has not been exploited in any non-trival way, 
presumably because it is not an invariance of the classical action 
nor is it a gauge symmetry. Nevertheless, we develop a technique 
for exploring the consequences of the symmetry leading to a 
conserved vector current and charge. The current and charge 
are identically zero in the free field case and only contribute 
in a background such as a electromagnetic or gravitational field. 
The current can couple spin-$\frac{3}{2}$ fields to vector and 
scalar fields and may have important consequences in intermediate 
energy hadron physics as well as linearized supergravity. 
The consistency problem which plagues higher spin field theories 
is then discussed and and some ideas regarding the possiblity of 
solutions are presented.

\end{abstract}

\section{Introduction}

Theories of interacting high spin fields\footnote{By `high spin' we mean
particles of spin $\geq \frac{3}{2}$.} have been a subject of 
considerable interest for many years. This is partly due to the many
particles with spin $\geq \frac{3}{2}$ seen in accelerator laboratories 
and also because there is currently no general field theory description 
which is relativistic, interacting and also free of 
inconsistencies\footnote{See section \ref{consistency} below.}.
Over the years one interacting theory after another have been shown to 
be inconsistent leading many to suggest that all higher spin fields 
must be composite. 
On the other hand, higher spin elementary particles, such as the 
gravitino, 
play an important role in supersymmetry, which itself represents a
fundamental building block of many modern unification schemes. 
Thus we would like to remain hopeful that a solution to the consistency 
problems can be found within point particle field theory.
Perhaps our interpretation regarding the physical degrees of
freedom is misguided or perhaps, as is our present concern, we have 
neglected symmetries or other aspects of nature that should 
be included. The hope being that if all of the symmetries are properly 
included, the result will be a consistent theory. 
That this hope is reasonable is exemplified by the fact that consistent 
solutions have already been found in restricted scenarios with curved 
backgrounds, cosmological constant tuning and Planck scale masses 
\cite{madore1975,deser1977,rindani1986,rindani1991,deser2001,deser2001-b}.

The consistency problems seem to exist for most interacting higher 
spin field theories and are a main concern of many theorists working 
in the field and so we will devote the final section of this paper to 
a discussion of the problem and touch on some possible consequences 
of the symmetry. 
Perhaps the ideas that we present will inspire some new angles of 
attack on the problem.

The main goal of this paper is a review of the $d$-dimensional theory 
of interacting Rarita-Schwinger fields and an exploration of the symmetries. 
Our hope is that the general expressions and the new interactions that we
present here will be of use in formulating effective theories of interacting 
hadrons as well as work involving the massive gravitinos of supergravity 
such as, for example, the AdS/CFT correspondence 
\cite{volovich1998,rashkov1999,koshelev1998,matlock1999}.
Perhaps the AdS/CFT results can be extended to the case wherein
the Rarita-Schwinger fields are not fixed at the start as non-interacting 
and onshell\footnote{We say non-interacting since onshell fields
are used and condition $\Gamma \cdot \psi = 0$ is incompatible with
the presence of additional non-gravitational interactions, for 
example a non-zero background electromagnetic field.}.

We begin in sections \ref{spincontent} and \ref{conditions}
by (re-)deriving the most general lagrangian and propagator for a 
Rarita-Schwinger spin-$\frac{3}{2}$ field\footnote{We follow the
standard procedure of using 4-dimensional terminology for spin even
when discussing fields in arbitrary dimensions.} using the method of 
Aurilia and Umezawa \cite{aurilia1969} extended to $d$-dimensions.

Since we are using the vector-spinor representation of spin-$\frac{3}{2}$,
we find the usual result that a lower spin content is retained
in the field in order to maintain the desirable properties of the action 
such as hermiticity, linearity in derivatives and non-singular behavior.
However, recently there have been other promising ideas where the
lower spin content is given a physical interpretation \cite{kaloshin2004} 
or where vector spinor description of spin-$\frac{3}{2}$ is replaced by a 
pure spin-$\frac{3}{2}$ field 
\cite{kirchbach2001,ahluwalia1992,ahluwalia1993,kirchbach2002}. 

Considering the lower spin components as unphysical, as we do here,
leads to a non-unique action depending on an arbitrary complex 
parameter measuring the lower spin content of the theory. 
Various choices of the parameter are seen to reduce the general expression to 
the spin-$\frac{3}{2}$ actions found in the literature.  
We formulate the equations in $d$ spacetime dimensions in anticipation 
of diverse applications from effective theories of hadronic interactions 
involving the spin-$\frac{3}{2}$ baryons to applications in arbitrary 
dimensional supergravity theories.
For example, both the composite $\Delta(1232)$ resonance found in low 
and intermediate energy nucleon scattering experiments and the gravitino 
of N-extended supersymmetry after spontaneous symmetry breaking are thought
to be described by the massive, spin-$\frac{3}{2}$, Rarita-Schwinger field 
that we study here.

In section \ref{group} we examine the properties of the so-called
`point' or `contact' transformations. These form a non-unitary group 
of transformations of the fields which shifts the parameter, amounting 
to a sort of rotation among the spin-$\frac{1}{2}$ degrees of freedom.
The path integral is seen to be invariant under point transformations 
which implies that physical correlation functions are invariant under 
a redefinition of the arbitrary parameter. That the parameter is arbitrary 
is well known and this has caused many authors to simply fix it to a 
convenient value. Unfortunately, this has served to hide some of the 
freedom of the theory. 
We restore the explicit parameter dependence and, in sections 
\ref{interactions} and \ref{implications}, we derive and examine new 
conserved currents resulting from the the symmetry. 

Finally, in section \ref{consistency} we discuss the consistency problems
and point out a few ideas of how the conserved charge found in
section \ref{implications} might be useful in that context.
The analysis we have used should also be generalizable to higher 
spins as well whenever the theory contains auxiliary fields of lower spin 
and has a similar symmetry group involving them. 

\section{Spin content of the Rarita-Schwinger field}
\label{spincontent}

In this section we give a decomposition of the Rarita-Schwinger field 
into separate spin blocks and derive some general formulas and identities 
that will be needed later. 
The result of this and the following section is the expression for the 
most general free lagrangian.
The reader only interested in the result may want to turn 
immediately to equation (\ref{action1}) or (\ref{Action1}) below.

A commonly used formulation of the spin-$\frac{3}{2}$ field is 
the vector-spinor representation\footnote{However there are other 
representations which describe spin-$\frac{3}{2}$ also, for example
the 3 spinor representation \cite{aurilia1969} and also the
$\left(\frac{3}{2}, 0\right) + \left(0, \frac{3}{2}\right)$ representation
of \cite{ahluwalia1992,ahluwalia1993}.}
given by Rarita and Schwinger in 1941 \cite{rarita1941}.
The vector-spinor transforms under the Lorentz group 
as\footnote{See section 1.2.2 in van Nieuwenhuizen's supergravity 
review \cite{van1981}}
\begin{equation}
\label{spindecomp}
\left(\frac{1}{2},\frac{1}{2}\right) \otimes \left[ \left(\frac{1}{2}, 0
\right) \oplus \left( 0, \frac{1}{2} \right) \right] 
= \left(1, \frac{1}{2}\right) \oplus \left(\frac{1}{2}, 1 \right) 
\oplus \left( 0, \frac{1}{2} \right) \oplus \left(\frac{1}{2}, 0 \right)
\end{equation}
whereas the spin decomposition of the field {\it in the rest frame} 
\cite{kirchbach2002,kaloshin2004} is
\begin{equation}
\text{spin } \psi_\mu^A = \left( 1 + 0 \right) \otimes \frac{1}{2}
= \frac{3}{2} + \frac{1}{2} + \frac{1}{2}.
\end{equation}
The vector-spinor field thus contains two spin-$\frac{1}{2}$ 
components in addition to the physical spin-$\frac{3}{2}$ component. 
The decomposition of the spin-$\frac{3}{2}$ field that we will use is given
by choosing $\left( 0, \frac{1}{2} \right) = p_\mu \psi^\mu$, where
$p_\mu = i \partial_\mu$. The complimentary part is 
\begin{equation}
\left(1, \frac{1}{2}\right) = \left( g_{\mu \nu} - \frac{p_\mu p_\nu}{p^2} \right) \psi^\nu,
\end{equation}
which can then be written in terms of spin-$\frac{3}{2}$ and 
spin-$\frac{1}{2}$ projectors as
\begin{equation}
g_{\mu \nu} - \frac{p_\mu p_\nu}{p^2} 
= \left(P^{\frac{3}{2}}\right)_{\mu \nu} + \left(P^{\frac{1}{2}}_{11}\right)_{\mu \nu}.
\end{equation}
Defining $\left(P^{\frac{1}{2}}_{22}\right)_{\mu \nu} 
= \frac{p_\mu p_\nu}{p^2}$ we have an expansion of the identity
\begin{equation}
\label{expanse1}
g_{\mu \nu} = \left(P^{\frac{3}{2}}\right)_{\mu \nu} 
+ \left(P^{\frac{1}{2}}_{11}\right)_{\mu \nu} 
+ \left(P^{\frac{1}{2}}_{22}\right)_{\mu \nu}.
\end{equation}
Explicit expressions for the projectors can easily be found by 
contraction with $\gamma^\mu$ and $p^\mu$, but we will anticipate
further 
applications\footnote{For example one may want to set $d = 4 - \epsilon$ 
in dimensional regularization.} and generalize to $d$-dimensions. 

We define $d$-dimensional spin projection operators by requiring that 
they reduce properly to the usual 4 dimensional projection operators 
\cite{van1981,benmerrouche1989,bernard2003} and yet remain projections in 
$d$-dimensions satisfying the following orthogonality relations,
\begin{equation}
\label{orthog}
\left( P^I_{ij} \right)_{\mu \nu} \left( P^J_{kl} \right)^{\nu
\rho} = \delta^{IJ} \delta_{jk} \left( P^I_{il}
\right)^{\rho}_{\mu}, \quad I, J \in \{ 1/2,3/2 \}, \quad i,j,k,l \in \{1,2\}.
\end{equation}
The result is
\begin{equation}
\label{projectors}
\begin{split}
\left( P^{\frac{3}{2}} \right)_{\mu \nu} &= \frac{1}{p^2 \left(d -
1\right)} \left[ \left(d-1 \right) p^2 g_{\mu \nu} -
\left(d-2\right) p_\mu p_\nu - \nt{p}
\left( \gamma_\mu p_\nu - p_\mu \gamma_\nu \right) - \gamma_\mu
\gamma_\nu p^2 \right] \\
\left( P^{\frac{1}{2}} \right)_{\mu \nu} &= \frac{1}{p^2 \left(d -
1\right)} \left[ \left(d-2 \right) p_\mu p_\nu + \nt{p}
\left( \gamma_\mu p_\nu - p_\mu \gamma_\nu \right) + \gamma_\mu
\gamma_\nu p^2 \right] = P^{\frac{1}{2}}_{11} +  P^{\frac{1}{2}}_{22} \\
\left( P^{\frac{1}{2}}_{11} \right)_{\mu \nu} &= \frac{1}{p^2 \left(d -
1\right)} \left[ - p_\mu p_\nu + \nt{p}
\left( \gamma_\mu p_\nu - p_\mu \gamma_\nu \right) + \gamma_\mu
\gamma_\nu p^2 \right] \\
\left( P^{\frac{1}{2}}_{22} \right)_{\mu \nu} &=
\frac{p_\mu p_\nu}{p^2} \\
\left( P^{\frac{1}{2}}_{12} \right)_{\mu \nu} &= \frac{1}{p^2 \sqrt{d -
1}} \left[ p_\mu p_\nu - \nt{p} \gamma_\mu p_\nu \right] \\
\left( P^{\frac{1}{2}}_{21} \right)_{\mu \nu} &= \frac{1}{p^2 \sqrt{d -
1}} \left[ \nt{p} p_\mu \gamma_\nu - p_\mu p_\nu \right].
\end{split}
\end{equation}
From the definitions we see that the total spin-$\frac{1}{2}$ projection
operator, $P^{\frac{1}{2}}$, reduces to the sum of the individual projection 
operators for the two different spin-$\frac{1}{2}$ components of the 
Rarita-Schwinger field.
We also note the following convenient relations,
\begin{equation}
\label{relations}
\begin{split}
\left[ \nt{p}, \left( P^{\frac{1}{2}}_{JJ} \right)_{\mu \nu}
\right] &= \left[ \nt{p}, \left( P^{\frac{3}{2}} \right)_{\mu \nu}
\right] = \left\{ \nt{p}, \left( P^{\frac{1}{2}}_{12} \right)_{\mu \nu}
\right\} = \left\{ \nt{p}, \left( P^{\frac{1}{2}}_{21} \right)_{\mu \nu}
\right\} = 0, \\
\gamma^\mu \left( P^{\frac{3}{2}}\right)_{\mu \nu} &= 
\left( P^{\frac{3}{2}}\right)_{\mu \nu} \gamma^\nu = 
p^\mu \left( P^{\frac{1}{2}}_{12}\right)_{\mu \nu} = 
\left( P^{\frac{1}{2}}_{21}\right)_{\mu \nu} p^\nu = 0, \\
\gamma^\mu \left( P^{\frac{1}{2}}\right)_{\mu \nu} &= 
\left( P^{\frac{1}{2}}\right)_{\nu \mu} \gamma^\mu = \gamma_\nu, \\
\gamma^\mu \left( P^{\frac{1}{2}}_{11}\right)_{\mu \nu} &= 
\left( P^{\frac{1}{2}}_{11}\right)_{\nu \mu} \gamma^\mu =
\gamma_\nu - \frac{\nt{p} p_\nu}{p^2}, \\
\gamma^\mu \left( P^{\frac{1}{2}}_{22}\right)_{\mu \nu} &= 
\left( P^{\frac{1}{2}}_{22}\right)_{\nu \mu} \gamma^\mu =
\frac{\nt{p} p_\nu}{p^2}.
\end{split}
\end{equation}
These projection operators can now be used to derive the most
general vector-spinor lagrangian and propagator in a flat\footnote{Curved 
space results can then be found by the usual technique of introducing 
vielbeins and a spin connection.} background spacetime.

\section{The free spin-$\frac{3}{2}$ lagrangian}
\label{conditions}

The free lagrangian for the spin-$\frac{3}{2}$ field can be written as
\begin{equation}
\label{freelagrangian}
\mathcal{L} = \ol{\psi}^\alpha \Lambda_{\alpha \beta} \psi^\beta,
\end{equation}
where $\Lambda_{\alpha \beta}$ is an operator and $\psi^\beta$ is a 
vector-spinor field with suppressed spin 
index\footnote{See \cite{wetterich1983} and references therein 
for properties of spinors in $d$ dimensions, the $d$-dimensional 
Lorentz group and the group of $d$-dimensional general coordinate 
transformations.}.
Using the projectors given in the
previous section we will construct the most general operator 
$\Lambda_{\alpha \beta}$ subject to the following four conditions
\cite{rarita1941,aurilia1969,moldauer1956,nath1971}:
\begin{enumerate}
\item{The Euler-Lagrange equations derived from the free action should give 
the local Rarita-Schwinger equations for a spin-$\frac{3}{2}$
particle. These are a Dirac equation for each of the vector components as 
well as supplementary conditions to remove the lower spin degrees of freedom:
\begin{equation}
\label{raritaschwinger}
\begin{split}
\left( i \nt{\partial} - m \right) \psi^\mu &= 0, \\
\gamma_\mu \psi^\mu &= 0.
\end{split}
\end{equation}
}
\item{The lagrangian should be non-singular in the limit $p \rightarrow 0$.
In particular, we would like the pole of the propagator to occur at the 
mass of the particle.}
\item{The lagrangian should be linear in derivatives as it describes 
a fermionic field.}
\item{The operator $\gamma^0 \Lambda_{\alpha \beta}$ should be hermitian:
\begin{equation}
\label{hermitian}
\gamma^0 \left(\Lambda_{\alpha \beta}\right)^\dagger \gamma^0 = \Lambda_{\beta \alpha}.
\end{equation}
}
\end{enumerate}
A consequence of equation (\ref{raritaschwinger}) in condition 1 is the 
condition
\begin{equation}
\partial_\mu \psi^\mu = 0,
\end{equation}
as can be seen by multiplying the first equation in 
(\ref{raritaschwinger}) 
on the left by $\gamma_\mu$ and using the second 
equation. 
The condition $\gamma_\mu^{AB} \psi^\mu_{B} = 0$ (where we now explicitly 
write the spinor indices $A,B$) represents a constraint equation for each 
value of the spin index $A$ whereas the condition $\partial_\mu \psi^\mu_B = 0$ is 
an equation of motion for the spinor components $\psi^0_B$. 
However, the Dirac equation (\ref{raritaschwinger}) also gives an equation of motion 
for the same spinor components and when taken together, these result in another 
set of constraints.
In four spacetime dimensions, these two sets of equations each constitute 
four constraints\footnote{This may not be true in the interacting
theory where the interaction may cause a reduction in the number of 
constraint equations resulting in an increase in the number of degrees of freedom
and leading to inconsistencies. This will be discussed further in
section \ref{consistency}.} and serve to remove 8 components of the 16 
component vector-spinor $\psi^\mu_A$, leaving $2(2 s + 1) = 8$ physical 
degrees of freedom as required for a massive spin $s = \frac{3}{2}$ 
particle\footnote{The restriction of condition 2 is perhaps not necessary, both the 
photon and the gluon propagators also have extra singular terms before gauge 
fixing and indeed some of the popular spin-$3/2$ actions do not satisfy 
this condition. However, in our case we have a theory describing a 
{\it massive} field and so extra terms which are singular as 
$p \rightarrow 0$ give singularities which are not at the physical 
mass unlike the case of the photon or the gluon.}.

The most general expression for the operator $\Lambda_{\alpha \beta}$
which obeys condition 1 is given by a combination of a
spin-$\frac{3}{2}$ part plus an arbitrary amount of spin-$\frac{1}{2}$,
\begin{equation}
\label{operator}
\Lambda_{\alpha \beta} = \left( \nt{p} - m \right)
P^{\frac{3}{2}}_{\alpha \beta}
+ \frac{a_1}{d} m \left( P^{\frac{1}{2}}_{11}\right)_{\alpha \beta} 
+ \frac{a_2}{d} m \left( P^{\frac{1}{2}}_{22}\right)_{\alpha \beta},
\end{equation}
where we have included factors of $m$ and $1/d$ in the two spin-$\frac{1}{2}$ 
terms for later convenience. 
The quantities $P^{\frac{3}{2}}_{\alpha \beta}$, $P^{\frac{1}{2}}_{11}$
and $P^{\frac{1}{2}}_{22}$ are the projection operators defined previously 
which, respectively, project onto the spin-$\frac{3}{2}$ part and the 
two spin-$\frac{1}{2}$ parts of $\psi^\beta$.
The Euler-Lagrange equations $\Lambda_{\alpha \beta} \psi^\beta =
0$ separate because of these projection operators into the Rarita-Schwinger 
equation plus supplementary conditions as required by condition 1. 
The equation also contains two complex constants $a_1$ and $a_2$ which are 
arbitrary at the moment. One can verify, using the projectors (\ref{projectors}) 
that the operator (\ref{operator}) satisfies condition 1 for any choice of 
these constants and so we are free to fix them to whatever values are convenient.
We will fix them by requiring the lagrangian to satisfy conditions 2 and 3.

Before proceeding with the other conditions let us take a moment to examine the general 
action given by (\ref{operator}). It contains possible gauge invariances 
which are already apparent. The field equations are seen to be
\begin{equation}
\label{fldeqn}
\left( \nt{p} - m \right)
P^{\frac{3}{2}}_{\alpha \beta} \psi^\beta
+ \frac{a_1}{d} m \left( P^{\frac{1}{2}}_{11}\right)_{\alpha \beta}  \psi^\beta
+ \frac{a_2}{d} m \left( P^{\frac{1}{2}}_{22}\right)_{\alpha \beta} \psi^\beta =
0
\end{equation}
and using (\ref{orthog}) and (\ref{relations}) we notice that, if we could 
set $a_1 = a_2 = 0$, the equation would be invariant under the two 
variations 
$\delta \psi^\beta = \left(P^{\frac{1}{2}}_{IJ}\right)^{\beta \lambda}
\chi_\lambda$
for arbitrary spinor $\chi_\lambda$. However, we will see that the remaining conditions 
require the parameters $a_1$ and $a_2$ to be related to each other in such a way 
that the vanishing of one implies the other is non-zero 
and so both parameters cannot be set to zero at the same time. 
If we set one of them to zero, i.e. $a_I \rightarrow 0$ for $I \in 1,2$, the equations 
of motion become invariant under 
$\delta \psi^\beta = \left(P^{\frac{1}{2}}_{II}\right)^{\beta \lambda}
\chi_\lambda$.
For $a_I = a_2$ this is the same as 
$\delta \psi^\beta = \partial^\beta \epsilon$ as can 
be seen by inserting the explicit form of the projector from (\ref{projectors}). 
We will see that our remaining conditions break this symmetry such 
that the gauge invariance is lost except in the massless limit (see equation
\ref{gaugeinvariance}).
We will examine the lagrangian written in terms of projection 
operators again at the end of this section after we have imposed the remaining 
conditions 2 -- 4.

Returning to our conditions, we see that the operator (\ref{operator}) as it is 
written, does not obey condition 2 since it is singular in the limit $p \rightarrow 0$. 
To remedy this, we use the method of Aurilia and Umezawa \cite{aurilia1969} 
and shift the spin-$\frac{1}{2}$ components to form a new operator as
\begin{equation}
\label{newoperator}
\tilde{\Lambda}_{\alpha \beta} = \left( \eta_2 \eta_1
\right)_{\alpha}^{\lambda} \Lambda_{\lambda \beta},
\end{equation}
with $\eta_1$ and $\eta_2$ given by
\begin{equation}
\label{transforms}
\begin{split}
\eta_1^{\mu \nu} &= g^{\mu \nu} + \sqrt{d-1}\left[ \frac{g_1}{m} \nt{p} +
g_2 \right] \left(P^{\frac{1}{2}}_{12}\right)^{\mu \nu}, \\
\eta_2^{\mu \nu} &= g^{\mu \nu} + \sqrt{d-1}\left[ \frac{f_1}{m} \nt{p} +
f_2 \right] \left(P^{\frac{1}{2}}_{21}\right)^{\mu \nu}.
\end{split}
\end{equation}
The $f_1, g_1, f_2, g_2$ are new constants that we will
fix by requiring the singular terms to vanish. The transformation 
(\ref{newoperator}) is allowed since we are only altering the coefficients 
of the separate spin-$\frac{1}{2}$ projection operators in (\ref{operator}) 
and thus the equations of motion will still separate properly and condition 1 
is still obeyed.

Substituting (\ref{transforms}) into (\ref{newoperator}) and
making use of the relations (\ref{relations}) we find that
the singular terms will vanish if the constants $f_1, g_1, f_2,
g_2, a_1$ and $a_2$ satisfy the following relations,
\begin{equation}
\label{params}
\begin{split}
\frac{a_2}{d} &= - \frac{d}{(d-1)} \left(1 + \frac{1}{a_1}
\right), \\
f_2 &= \frac{a_1 + d}{(d-1) a_1} = -\left(1 +
\frac{a_2}{d}\right), \\
g_2 &= \frac{a_1 + d}{(d-1) a_2} =
- \frac{a_1}{d} \left( 1 + \frac{d}{a_2} \right), \\
a_1 f_1 &= a_2 g_1 + \frac{(d-2)a_1}{(d-1)}.
\end{split}
\end{equation}
The resulting non-singular operator is then
\begin{equation}
\label{nonsingularoperator}
\begin{split}
\tilde{\Lambda}^{\alpha \beta} = \left( \nt{p} - m \right)
g^{\alpha \beta} &- \left[ \frac{a_1}{d} f_1 + \frac{1}{(d-1)} \right]
\left( \gamma^\alpha p^\beta - p^\alpha \gamma^\beta \right) \\
&+ \frac{(d-2)}{(d-1)} \frac{a_1}{d} \gamma^\alpha p^\beta 
- \frac{1}{(d-1)} \nt{p} \gamma^\alpha \gamma^\beta -
(d-1)\frac{a_2}{d} f_1 g_1
\frac{p^\alpha p^\beta}{m}.
\end{split}
\end{equation}
This operator now satisfies conditions 1 and 2, and it is written in 
terms of two parameters, $f_1$ and $a_1$ (since the relations 
(\ref{params}) can be used to write $a_2$ and $g_1$ in terms of 
$a_1$ and $f_1$).

We would like the field $\psi^\mu$ and its hermitian conjugate 
to appear symmetrically in the lagrangian, so that 
the variation of $\psi^\mu$ and that of $\ol{\psi}^\mu$ both give 
the field equations. To achieve this, we apply another shift to 
(\ref{nonsingularoperator}) via the
transformation\footnote{See for example equation (4.2) in 
Johnson and Sudarshan \cite{johnson1961} or equation (15) in van
Nieuwenhuizen's supergravity review \cite{van1981}.} 
\begin{equation}
\label{pointtrans}
\theta^{\mu \nu}(k) = g^{\mu \nu} + \frac{k}{d} \gamma^\mu \gamma^\nu.
\end{equation}
Again, we see that the equations of motion remain unaffected
(since $\gamma^\nu \psi_\nu = 0$ onshell) and thus the resulting operator 
will still satisfy conditions 1 and 2. 
The purpose of this transformation is that now our operator contains 
a new constant, $k$, which we can use in satisfying the other conditions. 
Our general operator is then
\begin{equation}
\label{blah}
\begin{split}
&\Lambda^{\alpha \beta} = \theta^{\alpha \lambda}(k)
\tilde{\Lambda}_{\lambda}^{\beta} \\
&= \left( \nt{p} - m \right) g^{\alpha \beta} 
- \left[ \frac{a_2}{d} g_1 + \frac{1}{(d-1)} \right] 
\left( \gamma^\alpha p^\beta - p^\alpha \gamma^\beta \right) 
- \left[ \frac{k a_1 f_1}{d^2} + \frac{1}{(d-1)} \right] 
\nt{p} \gamma^\alpha \gamma^\beta \\
&+ \left[ 1 + \frac{a_1}{d} + k \left(
\frac{1}{d} + \frac{a_1}{d} \right) \right] \frac{m \gamma^\alpha
\gamma^\beta}{(d-1)} - \frac{(d-1) a_2 f_1 g_1}{m d} \left[ \left(
1 + \frac{2 k}{d} \right) p^\alpha p^\beta - 
\frac{k}{d} \nt{p} \gamma^\alpha p^\beta \right] \\
&+ \frac{k}{d} \left[ \frac{(d-2)}{(d-1)} - a_2 g_1 \right] 
\gamma^\alpha p^\beta + \left[ \frac{2k}{d^2} a_1 f_1 +
\frac{(d-2)}{d (d-1)} a_1 \right] p^\alpha \gamma^\beta.
\end{split}
\end{equation}

Condition 3 requires the lagrangian to be linear in derivatives. 
Inspection of (\ref{blah}) reveals that 
the terms quadratic in $p^\alpha$ can be explicitly cancelled by setting 
$g_1 = 0$ or $f_1 = 0$. This gives two paths to follow: 
setting $g_1 = 0$ gives
\begin{equation}
\label{op1}
\begin{split}
\Lambda^{\alpha \beta}_1 &= \left( \nt{p} - m \right) g^{\alpha \beta} 
- \frac{1}{(d-1)} \left( \gamma^\alpha p^\beta - p^\alpha \gamma^\beta \right) 
- \frac{1}{(d-1)} \left[ \frac{k a_1 (d-2)}{d^2} + 1 \right] 
\nt{p} \gamma^\alpha \gamma^\beta \\
&+ \left[ 1 + \frac{a_1}{d} + \frac{k}{d} \left( 1 + a_1 \right) \right] 
\frac{m \gamma^\alpha \gamma^\beta}{(d-1)} 
+ \frac{k(d-2)}{d(d-1)} \gamma^\alpha p^\beta 
+ \frac{a_1(d-2)}{d(d-1)}\left[ \frac{2k}{d} + 1 \right] p^\alpha \gamma^\beta,
\end{split}
\end{equation}
whereas setting $f_1 = 0$ gives
\begin{equation}
\label{op2}
\begin{split}
\Lambda^{\alpha \beta}_2 = \left( \nt{p} - m \right) g^{\alpha \beta} 
&- \frac{1}{(d-1)} \left( \gamma^\alpha p^\beta - p^\alpha \gamma^\beta \right) 
- \frac{1}{(d-1)} \nt{p} \gamma^\alpha \gamma^\beta \\
&+ \left[ \frac{\ol{a}_1}{d} + 1 \right] \frac{m \gamma^\alpha \gamma^\beta}{(d-1)} 
+ \frac{\ol{a}_1 (d-2)}{d(d-1)} \gamma^\alpha p^\beta.
\end{split}
\end{equation}
where we have defined $\ol{a}_1 = a_1 + k \left( 1 + a_1 \right)$.

We now have two operators, the first (\ref{op1}) depending on
two parameters, $a_1$ and $k$, and the second (\ref{op2}) depending
on only one, $\ol{a}_1$. Both operators satisfy conditions 1
through 3. The last condition to impose on our operators is
condition 4, i.e. that the operators be hermitian.
Imposing the hermitian requirement (\ref{hermitian}) on the first
operator (\ref{op1}) fixes a relation between $a_1$ and $k$
\begin{equation}
\label{cond1}
a_1^* = k
\end{equation}
thus reducing the number of parameters to one.
The lagrangian (\ref{op1}) is then
\begin{equation}
\label{lagrangian}
\begin{split}
\Lambda^{\alpha \beta}_1 = \left( \nt{p} - m \right)& g^{\alpha \beta} 
- \frac{1}{(d-1)} \left( \gamma^\alpha p^\beta + p^\alpha \gamma^\beta \right) 
+ \frac{(d-2)}{d^2(d-1)} \left[ |a_1|^2 + \frac{d^2}{(d-2)} \right] 
\gamma^\alpha \nt{p} \gamma^\beta \\
&+ \left[ d + a_1 + a_1^* + |a_1|^2 \right] 
\frac{m \gamma^\alpha \gamma^\beta}{d(d-1)} 
+ \frac{(d-2)}{d(d-1)} \left( a_1^* \gamma^\alpha p^\beta 
+ a_1 p^\alpha \gamma^\beta \right).
\end{split}
\end{equation}
If we were to restrict $a_1$ to be real, and define 
the number $A$ in terms of $a_1$ as
\begin{equation}
\label{defna}
a_1^* = a_1 = \frac{d (d-1)}{d-2}A + \frac{d}{d-2}
\end{equation}
then $\Lambda^{\alpha \beta}_1$ would become
\begin{equation}
\label{benmer}
\begin{split}
\Lambda^{\alpha \beta}_{1A} = \left( \nt{p} - m \right) g^{\alpha \beta} 
&+ A \left( \gamma^\alpha p^\beta + p^\alpha \gamma^\beta \right) 
+ \frac{1}{(d-2)} \left[ (d-1)A^2 + 2A + 1\right] 
\gamma^\alpha \nt{p} \gamma^\beta \\
&+ \frac{m \gamma^\alpha \gamma^\beta}{(d-2)^2} 
\left[ d(d-1)A^2 + 4(d-1)A + d \right],
\end{split}
\end{equation}
which is the $d$-dimensional form of a common expression found in the 
literature \cite{benmerrouche1989,moldauer1956,haberzettl1998,munczek1967}.

Imposing the hermitian requirement on our second operator (\ref{op2}), 
gives the condition
\begin{equation}
\label{cond2}
\begin{split}
\ol{a}_1 &= 0 \\
\Rightarrow a_1 &= \frac{-k}{k+1},
\end{split}
\end{equation}
and the second lagrangian (\ref{op2}) becomes
\begin{equation}
\label{rarita}
\Lambda^{\alpha \beta}_2 = \left( \nt{p} - m \right) g^{\alpha \beta} 
- \frac{1}{(d-1)} \left( \gamma^\alpha p^\beta + p^\alpha \gamma^\beta \right) 
+ \frac{1}{(d-1)} \gamma^\alpha \left(\nt{p} + m \right) \gamma^\beta 
\equiv \Lambda^{\alpha \beta}_{\text{RS}} 
\end{equation}
which is the $d$-dimensional Rarita-Schwinger lagrangian
\cite{rarita1941} and has no arbitrary parameters.

We have thus found the most general set of $d$-dimensional 
lagrangians for a spin-$\frac{3}{2}$ field which satisfy our four
conditions. These are given by (\ref{lagrangian}) which depends on a 
complex parameter, and (\ref{rarita}) which is the usual 
Rarita-Schwinger lagrangian and has no arbitrary parameters.
In fact, the second lagrangian (\ref{rarita}) corresponds to the case $a_1
\rightarrow 0$ of the first lagrangian (\ref{lagrangian}). 
So the most general expression for the set of 
operators is given by the single expression (\ref{lagrangian}) with 
the Rarita-Schwinger operator corresponding to the particular value 
$a_1 = 0$ of the arbitrary parameter. 
It will be shown in section \ref{group} that this
general operator can be written in a very simple manner as a 
transformation of the Rarita-Schwinger lagrangian with $a_1$ being the 
parameter of the transformation -- a fact already noticed by Freedman and 
van Nieuwenhuizen \cite{freedman1976} in 1976. 

The propagator, $S^{\alpha \beta}$, for the spin-$\frac{3}{2}$ field 
is the inverse of the quadratic operator in the lagrangian
\begin{equation}
\label{invexpr}
S_{\alpha \beta} \Lambda^{\beta \lambda} = \delta_\alpha^\lambda \; .
\end{equation}
The definition of the projection operators
(\ref{projectors}) gives the equation
\begin{equation}
\left( p^2 - m^2 \right) S_{\alpha \beta} = 
\left(\nt{p} + m \right) P^{\frac{3}{2}}_{\alpha \beta}
+ \frac{ \left(p^2 - m^2 \right) d}{m} \left[
\frac{\left(P^{\frac{1}{2}}_{11}\right)_{\alpha \beta}}{a_1} 
+ \frac{\left(P^{\frac{1}{2}}_{22}\right)_{\alpha \beta}}{a_2} \right],
\end{equation}
which can be solved in a similar way as we have done for the lagrangian, but
we will skip this lengthy calculation and simply quote the result
(\ref{propagator1}) below.

To summarize, the most general lagrangian for the spin-$\frac{3}{2}$ field 
consistent with our four conditions\footnote{See also the equivalent, but much
simpler, expression given by (\ref{Action1}) and (\ref{action2}) in section 
\ref{implications} below.} is
\begin{equation}
\label{action1}
\begin{split}
&\mathcal{L} 
= \ol{\psi}^\alpha \Biggl\{
\left( \nt{p} - m \right) g_{\alpha \beta} 
- \frac{1}{(d-1)} \left( \gamma_\alpha p_\beta + p_\alpha \gamma_\beta \right) 
+ \frac{1}{(d-1)} \gamma_\alpha \left(\nt{p} + m \right) \gamma_\beta \\
+ &\frac{(d-2)}{d^2(d-1)} |a|^2 
\gamma_\alpha \nt{p} \gamma_\beta + \left[a + a^* + |a|^2 \right] 
\frac{m \gamma_\alpha \gamma_\beta}{d(d-1)} 
+ \frac{(d-2)}{d(d-1)} \left( a^* \gamma_\alpha p_\beta 
+ a p_\alpha \gamma_\beta \right) \Biggr\} \psi^\beta, \\
&\quad = \ol{\psi}^\alpha \theta_{\alpha \mu}(a^*)
\; \Lambda^{\mu \nu}_{\text{RS}} \; \theta_{\nu \beta}(a) \psi^\beta,
\end{split}
\end{equation}
where $a$ is a complex parameter with the restriction $a \neq -1$. 
Notice that we have dropped the index on $a_1$ and will henceforth write
it as simply $a$ since only this single parameter remains. 
The last line in (\ref{action1}) is a more compact way of writing 
the expression as will be shown in the next section.
The propagator is
\begin{equation}
\label{propagator1}
\begin{split}
S^{\alpha \beta} = 
&\frac{1}{\left(\nt{p} - m \right)} \left[ g^{\alpha \beta}
- \frac{\gamma^\alpha \gamma^\beta}{(d-1)} 
- \frac{\left(\gamma^\alpha p^\beta - p^\alpha \gamma^\beta \right)}{(d-1)m}
- \frac{(d-2)}{(d-1)} \frac{p^\alpha p^\beta}{m^2} \right] \\
+ &\frac{(d-2)}{m^2(d-1)} \left( 
h^* \gamma^\alpha p^\beta + h p^\alpha
\gamma^\beta - |h|^2 \gamma^\alpha \nt{p}
\gamma^\beta \right) 
+ \frac{|h|^2 d - h - h^*}{(d-2)m} \gamma^\alpha \gamma^\beta, \\
= &\; \theta_{\alpha}^{\; \mu}(\tilde{a}) \; S_{\mu \nu}^{RS} \; \theta^{\nu}_{\;
\beta}(\tilde{a}^*),
\end{split}
\end{equation}
where $h = \frac{d + a^*}{d(1+a^*)}$.
Again, the last line in (\ref{propagator1}) is a more compact way of 
writing the expression as will be shown in the next section.
The inverse parameter $\tilde{a}^*$ is related to $h$ in our expression as
$\tilde{a}^* = \frac{d(1-h)}{1-d}$.
With real parameter $A$ defined by (\ref{defna}) we get a propagator 
often used \cite{benmerrouche1989,moldauer1956,nath1971,haberzettl1998,amiri1992},
\begin{equation}
\begin{split}
S^{\alpha \beta} = 
\frac{\left(\nt{p} + m \right)}{\left(p^2 - m^2 \right)} &\left[ g^{\alpha \beta}
- \frac{1}{(d-1)} \gamma^\alpha \gamma^\beta - \frac{1}{(d-1)m}
\left(\gamma^\alpha p^\beta - p^\alpha \gamma^\beta \right) 
- \frac{(d-2)}{(d-1)} \frac{p^\alpha p^\beta}{m^2} \right] \\
&+ \frac{A + 1}{m^2(Ad + 2)} \Biggl\{ 
\left[ \frac{d-4 - d A}{(d-2)(d A + 2)}\right] 
m \gamma^\alpha \gamma^\beta \\
&\quad + \frac{(d-2)}{(d-1)} \left(\gamma^\alpha p^\beta + p^\alpha
\gamma^\beta \right) 
- \frac{(d-2)(A + 1)}{(d-1) (dA + 2)} \gamma^\alpha \nt{p} \gamma^\beta 
\Biggr\},
\end{split}
\end{equation}
and setting $A = -1$ gives another common expression
\cite{bernard2003,dejong1992,pascalutsa1999,2pascalutsa2003,pascalutsa2003}.
The Rarita-Schwinger lagrangian and propagator \cite{rarita1941} 
correspond to the limits $a \rightarrow 0$ and $h \rightarrow 1$ 
whereas those 
of \cite{van1981,bernard2003,dejong1992,pascalutsa1999,pascalutsa2003} 
correspond to $a = -d$ (or $h = 0$). The latter expression is 
often mistakenly cited as the original Rarita-Schwinger action in 
the supergravity literature. 

Let us now take a moment to discuss the meaning of the arbitrary 
parameter $a$. The vector-spinor field contains auxiliary lower spin components 
which are necessary in maintaining Lorentz covariance in our formalism. 
This is a generic feature of the treatment of higher spin fields in 
relativistic lagrangian field theory with constraints. 
In order to get a theory which satisfies our general conditions
for the lagrangian, we introduced 
several arbitrary parameters. These parameters were not fully fixed by our 
conditions and instead we were left with one parameter, $a$, remaining unfixed.
We can see in (\ref{operator}) that $a_1$ and $a_2$ measure 
the relative strengths of the $P_{11}$ and $P_{22}$ parts of the field. 
Since $a_1$ and $a_2$ are inversely related by (\ref{params})
increasing $a$ corresponds to increasing the $P_{11}$
part and decreasing the $P_{22}$ part of the operator. 
Thus the parameter $a$ is a measure of the proportion of the two
different auxiliary spin-$\frac{1}{2}$ components of the theory. 
The general conditions which we used in formulating the theory has
resulted in this proportionality and transforming the parameter 
is equivalent to changing this proportionality. 
There is no choice of parameter which will eliminate both 
spin-$\frac{1}{2}$ components simultaneously and 
hence our conditions have forced us to retain
lower spin components in our theory, the `amount' of each depending on 
the choice of parameter. 

We can now return to the subject of gauge invariances that we touched
upon previously and see how the invariances of equation (\ref{operator}) 
have been modified by our conditions. 
The projection operators can be used to re-write the general 
lagrangian operator as 
\begin{equation}
\label{gaugeinvariance}
\begin{split}
\Lambda_{\alpha \beta} &= \left( \nt{p} - m \right) P^{\frac{3}{2}}_{\alpha \beta} \\
&+ \left[ \frac{(d-2)}{(d-1)} |a|^2 \left( \nt{p} -
m \right) - m \biggl( a^* b + b^* a \biggr) 
\right] \frac{(d-1)}{d^2} \left( P^{\frac{1}{2}}_{11}\right)_{\alpha \beta} \\
&+ \left[ (d-2) a^* b \left( \nt{p}
- m \right) - m \left( \frac{a^* (b+d)}{d-1} - |b|^2 \right) 
\right] \frac{(d-1)^\frac{3}{2}}{d^2} \left( P^{\frac{1}{2}}_{12}\right)_{\alpha \beta} \\
&+ \left[(d-2) |b|^2 \left( \nt{p} + m
\right) + m \biggl( (b^* + d)b + b^* (b+d) \biggr) 
\right] \frac{(d-1)}{d^2} \left( P^{\frac{1}{2}}_{22}\right)_{\alpha \beta} \\
&- \left[ (d-2) a b^* \left(
\nt{p} + m \right) + m \left( \frac{(b^*+d) a}{d-1} - |b|^2 \right) 
\right] \frac{(d-1)^\frac{3}{2}}{d^2} \left( P^{\frac{1}{2}}_{21}\right)_{\alpha \beta},
\end{split}
\end{equation}
where we have defined $b = \frac{a+d}{(1-d)}$.
This is the analogous equation to (\ref{operator}) except that the action 
now satisfies our four conditions and so the symmetry of (\ref{operator}) 
has been reduced.

The free equations of motion $\Lambda_{\alpha \beta} \psi^\beta = 0$ are 
invariant under $\psi^\beta \rightarrow \psi^\beta + \delta \psi^\beta$ 
whenever $\delta \psi^\beta$ is annihilated by $\Lambda_{\alpha \beta}$.
Using (\ref{gaugeinvariance}) along with (\ref{orthog}) we see that invariances 
occur only when the field is massless $m=0$ and also either $a = 0$ or $b = 0$. 
The case $a=0$ corresponds to the massless Rarita-Schwinger action and gives 
the invariance $\delta \psi^\beta = \left( p^\beta - \gamma^\beta \nt{p} \right) \epsilon$
for $\epsilon$ an arbitrary spinor.
This can be written as $\delta \psi^\beta = \theta^{\beta \lambda}(-d) \;
\partial_\lambda \epsilon$
using the results of the next section.
The case $b = 0$ corresponds to $a = -d$ and is the gravitino part
of the linearized supergravity action. In this case, the projectors give 
the invariance $\delta \psi^\beta = \partial^\beta \epsilon$ which 
is the linearized version of the gauge invariance in the gravitino action
of supergravity\footnote{In curved space let 
$\partial_\beta \rightarrow D_\beta = \partial_\beta + \frac{1}{4} 
\omega_{\beta}^{\; ab} \gamma_{ab}$ where $a,b$ are flat indices, 
$\omega_{\beta}^{\; ab}$ is the spin connection and 
$\gamma_{ab} \equiv \frac{1}{2} [\gamma_a, \gamma_b]$.}.

At first glance it may seem that the gauge invariance in
the massless limit only occurs for two values of the parameter. 
In fact, these two invariances represent merely specific cases of 
the same general invariance\footnote{The author
would like to thank V. Pascalutsa for pointing this out.}.
We form the general gauge transformation by using the
group properties of the point transformations which we discuss in the
following sections \cite{pascalutsa1999}, for example, 
$\delta \psi^\alpha = \theta^{\alpha \beta} ({\tilde{b}}) \; \partial_\beta \epsilon$ 
is a gauge invariance for arbitrary parameter because the inverse transformation 
$\theta^{\alpha \beta}({\tilde{b}})$ effectively sets $b=0$, as
will soon become clear. 

\section{The point transformation group}
\label{group}

The transformation (\ref{pointtrans}) that we have used in the
previous section is sometimes referred to as a `point' or `contact' 
transformation in the literature. 
The Rarita-Schwinger equations (\ref{raritaschwinger}) tell
us that the {\it onshell} spin-$\frac{3}{2}$ field 
$\psi^\mu$ satisfies the following equation
\begin{equation}
\label{holds}
\theta^\alpha_{\; \mu}(a) \psi^\mu = \psi^\alpha 
\end{equation}
as can be seen by the form of the transformation
(\ref{pointtrans}): 
$\theta^{\mu \nu}(a) = g^{\mu \nu} + \frac{a}{d} \gamma^\mu \gamma^\nu$,
along with the onshell constraint $\gamma \cdot \psi = 0$.
When $a \neq -1$ these transformations form a group with
\begin{equation}
\begin{split}
\theta^{\mu \nu}(a) \theta_{\nu \rho}(b) 
&= \theta^{\mu}_{\; \rho}(a + b + ab) \equiv \theta^{\mu}_{\; \rho}(a \circ b), \\
\left(\theta^{\mu \nu}\right)^{-1}(a) &= 
\theta^{\mu \nu}(\frac{-a}{1+a}) \equiv \theta^{\mu \nu}(\tilde{a}),
\end{split}
\end{equation}
where we have defined the `circle' operation 
$a \circ b = a + b + ab$ and the inverse parameter $\tilde{a} = \frac{-a}{1+a}$.
The lagrangian is tautologically invariant under the transformation 
$\psi^\alpha \rightarrow \theta^\alpha_{\; \mu}(k) \psi^\mu$
if, in addition to transforming the fields, we also transform the 
parameter $a$ as
\begin{equation}
\label{halftrans}
a^\prime = \frac{a - k}{1 + k}.
\end{equation}
Our field transformation together with the transformation (\ref{halftrans}) of
the parameter $a$ are frequently called a point 
transformation\footnote{To compare with the notations of other authors, 
notice that in the case where $k$ and $a$ are real we can use (\ref{defna}) to
rewrite the transformation of the parameter $a$ in terms of $A$ as
$A \rightarrow \frac{A - 2k/d}{1 + k}$. Also note our factor of $\frac{1}{d}$ 
in the group law.}

It is worthwhile to notice that separate transformations from
the left and from the right act separately on $a$ and $a^*$. 
If we recall how we formed our lagrangian by applying
a transformation from the right and how the requirement
that the operator be hermitian fixed $k = a^*$ we can see that 
\begin{equation*}
\begin{split}
\theta_\mu^{\; \alpha}(k^*) \Lambda_{\alpha \beta}(a^*,a) &= 
\theta_\mu^{\; \sigma}(k^*) \theta_\sigma^{\; \alpha}(a^*) 
\Lambda_{\alpha \beta}(a), \\
&= \theta_\mu^{\; \alpha}(k^* \circ a^*) \Lambda_{\alpha \beta}(a) \\
&= \Lambda_{\alpha \beta}(k^* \circ a^*,a). 
\end{split}
\end{equation*}
We find the left hand transformations by taking the hermitian
conjugate,
\begin{equation*}
\begin{split}
\Lambda_{\mu \nu}(a^*,a) \theta^{\nu}_{\; \beta}(k) 
&= \Lambda_{\mu \beta}(a^*, a \circ k) \; .
\end{split}
\end{equation*}
Hence transforming $\Lambda_{\mu \nu}(a^*,a)$ from the right affects only 
$a$, whereas the transformation from the left affects only $a^*$.
We can use these left and right transformations, along
with the definition of the Rarita-Schwinger lagrangian: 
$\Lambda^{\alpha \beta}_{\text{RS}} 
= \Lambda^{\alpha \beta}(0,0)$, to write our most general lagrangian as
\begin{equation}
\label{general}
\Lambda^{\alpha \beta}(a^*,a)
= \Lambda^{\alpha \beta}(a^* \circ 0, 0 \circ a)
=\theta^\alpha_{\; \mu}(a^*) \; \Lambda^{\mu \nu}_{\text{RS}} \;
\theta^{\; \beta}_{\nu}(a) \; .
\end{equation}
Substituting this expression into the relation (\ref{invexpr}) we have
\begin{equation}
\begin{split}
S_{\alpha \beta} \; \theta^\beta_{\; \mu}(a^*) \; \Lambda^{\mu \nu}_{\text{RS}} \; 
\theta^{\; \lambda}_{\nu}(a) &= \delta_\alpha^\lambda \\
\Rightarrow \left[ \theta^{\delta \alpha}(a) \; S_{\alpha \beta} \;  
\theta^{\beta \mu}(a^*) \right] \Lambda_{\mu \nu}^{\text{RS}} &= \delta^{\delta}_{\; \nu}.
\end{split}
\end{equation}
and the expression in brackets in the last line must therefore equal the 
Rarita-Schwinger propagator. So the point transformation group properties 
have given us a convenient way of deriving the general propagator from the
parameterless RS propagator which much simpler than the method used in 
section \ref{conditions}, namely:
\begin{equation}
\label{general2}
\begin{split}
S_{\alpha \beta}^{RS} &= 
\theta_{\alpha}^{\; \mu}(a) \; S_{\mu \nu} \; \theta^{\nu}_{\; \beta}(a^*) \\
\Rightarrow S_{\alpha \beta} &= 
\theta_{\alpha}^{\; \mu} (\tilde{a}) \; S_{\mu \nu}^{RS} \; \theta^{\nu}_{\; \beta}(\tilde{a}^*).
\end{split}
\end{equation}

The point transformation becomes singular at the parameter value $a = -1$ 
as can be seen by the fact that $k \circ -1 = -1$ for any $k$, so that
\begin{equation}
\label{singular}
\theta^{\mu \nu}(-1) \theta_{\nu \lambda}(k) = \theta^\mu_{\; \lambda}(-1)
\quad \forall \; k. 
\end{equation}
In the singular case $a = a^* = -1$ the transformations of the
fields would no longer change the lower spin content of the
operator and this choice may appear quite attractive. The problem with
the choice $a = -1$ is that the operator in the lagrangian no
longer has an inverse \cite{benmerrouche1989} as can be seen by the 
form of $h$ and $h^*$ in the propagator (\ref{propagator1}).
Interestingly, the singular value, $a = -1$, generates the additive identity 
element (or `zero') of a ring defined by the following addition rule
\begin{equation}
\begin{split}
\theta_{\mu \nu}(a) + \theta_{\mu \nu}(b) &= \theta_{\mu \nu}(a + b + 1), \\
\theta_{\mu \nu}(a) - \theta_{\mu \nu}(b) &= \theta_{\mu \nu}(a - b - 1), 
\end{split}
\end{equation}
where the addition is defined {\it modulo zero},
the two-sided ideal generated by $\theta_{\mu \nu}(-1)$. 
It is easily shown that the multiplication is distributive over addition.

One can redefine the parameter in various ways to make the group more convenient. 
For example, by shifting the singular point of the parameter space 
to $- \infty$ by letting $a \rightarrow e^{\alpha} - 1$ the group 
becomes \cite{pascalutsa1999}
\begin{equation}
\label{lorentz}
\begin{split}
\theta^{\mu \nu}(\alpha) &= g^{\mu \nu} + \frac{e^{\alpha} -
1}{d} \gamma^\mu \gamma^\nu = e^{\frac{\alpha}{d} \gamma^\mu
\gamma^\nu} \\
\theta^{\mu \nu}(\alpha) \theta_{\nu \lambda}(\beta) 
&= \theta^\mu_\lambda(\alpha + \beta), \\
\left(\theta^{\mu \nu}\right)^{-1}(\alpha) &= \theta^{\mu \nu}(-\alpha).
\end{split}
\end{equation}
An even more convenient redefinition is so that the singular point is at 0. 
One has $a \rightarrow \alpha - 1$ and the ring is then defined by
\begin{eqnarray*}
\theta^{\mu \nu}(\alpha) =& g^{\mu \nu} + \frac{\alpha - 1}{d} \gamma^\mu \gamma^\nu, 
& \text{(definition)}  \\
\theta^{\mu \lambda}(\alpha) \theta_{\lambda}^{\; \; \nu}(\beta) 
=& \theta^{\mu \nu}(\alpha \beta), 
& \text{(multiplication)}  \\
\theta^{\mu \nu}(1) =& g^{\mu \nu}, & \text{(multiplicative identity)}  \\
\left(\theta^{\mu \nu}\right)^{-1}(\alpha) =& \theta^{\mu \nu}(\frac{1}{\alpha}), 
& \text{(multiplicative inverse)}  \\
\theta^{\mu \nu}(\alpha) + \theta^{\mu \nu}(\beta) =& \theta^{\mu \nu}(\alpha + \beta), 
& \text{(addition)}  \\
\theta^{\mu \nu}(0) =& g^{\mu \nu} - \frac{1}{d} \gamma^\mu \gamma^\nu, 
& \text{(additive identity)}  \\
\theta^{\mu \nu}(\alpha) - \theta^{\mu \nu}(\beta) =& \theta^{\mu \nu}(\alpha - \beta),
& \text{(additive inverse)}
\end{eqnarray*}
where the addition is again defined modulo the additive identity. 

\section{Interactions}
\label{interactions}

The path integral is invariant under a global point transformation of the fields
since the functional determinant is trivial and factors out of the integral to be 
cancelled out of the generating functional by the identical factor in the denominator.
Hence there are no path integral anomalies and all physical correlation 
functions are independent of the parameter $a$. 
This is also true in the interacting theory, as we will discuss shortly, 
and so all physical Green functions are independent of
$a$ and it can be fixed to whatever value is convenient.
As was pointed out by Nath, {\it et al.} \cite{nath1971}, the
meaning of the invariance under point transformations is that the 
physical content of the theory does not depend on the parameter. 
However, the classical equations of motion in the presence of interaction
are {\it not} invariant under shifts of the parameter. 
This is due to the fact that the transformation is not unitary. 
This will be discussed further in the next section.

There has been some controversy about to how to include consistent
interactions involving spin-$\frac{3}{2}$ fields 
\cite{benmerrouche1989,nath1971,pascalutsa1999,hagen1982}.
We find the most logical way is to require the interaction terms to 
transform the same way under point transformations as the free action
as well as remaining consistent with the massless gauge invariance of
the free action. 
In other words, we require that a shift of the parameter must leave the form of 
the entire action unchanged. This ensures that the the path integral remains 
independent of the parameter and is the reason why many authors require the 
action to be invariant under a point transformation combined with a compensating 
parameter shift. This requirement means that we must have a factor
of $\theta_{\mu \nu}(a)$ for each $\psi^\nu$ field in the interaction lagrangian 
(and thus a factor of $\theta_{\mu \nu}(a^*)$ in the hermitian conjugate) so that 
a transformation $\psi^\alpha \rightarrow \theta^{\alpha}_{\; \beta}(k) \psi^\beta$ 
will shift $a \rightarrow a \circ k$ everywhere\footnote{Note that in order to be 
consistent with the massless gauge invariance of the theory, the
interaction must depend only on $d \psi$ rather than simply $\psi$ and 
also there should be no `offshell parameter' dependence. This is
discussed by Pascalutsa and Timmermans in \cite{pascalutsa1999}.}.

\begin{figure}[htb]
\begin{center}
\input{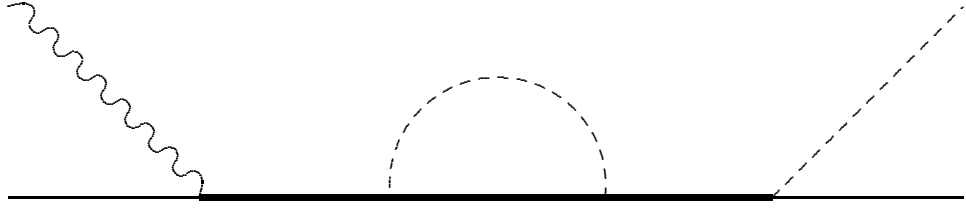}
\end{center}
\caption{A loop diagram found in resonant photo-pion production
exemplifying the cancellation of parameter dependence in Green functions.
The thin lines are spin-$\frac{1}{2}$ (nucleons), the dashed lines are spin-$0$
(pions), the wavy line is spin-$1$ (photon) and the thick line is the 
Rarita-Schwinger, spin-$\frac{3}{2}$ (e.g. $\Delta(1232)$) field.}
\label{pionfig}
\end{figure}

Expression (\ref{general2}) shows that, for interactions which depend on the 
parameter in the same way as the lagrangian operator in (\ref{general}), the 
inverse transformations attached to the RS propagator in (\ref{general2}) will 
cancel those of the interaction and will therefore lead to a theory whose 
Green functions are independent of the parameter.
We can see schematically that the correlation functions are invariant
under redefinitions of the parameter as follows.
In a theory satisfying our requirements the propagator and vertex can
be written, respectively, as (suppressing all indices for simplicity)
\begin{equation}
\begin{split}
S(a) &= \theta^{-1} S \theta^{-1}, \\
\Lambda(a) &= \theta \Lambda \theta,
\end{split}
\end{equation}
where the $a$ dependence on the right-hand-side is entirely contained in the 
$\theta$ factors: $S$ and $\Lambda$ are independent of $a$. 
This implies that products of these expressions found in 
Feynman diagrams will always reduce to the form
\begin{equation}
\begin{split}
G(a) &\sim \Lambda(a) S(a) \Lambda(a) \cdots \Lambda(a) S(a) \Lambda(a), \\
&= \theta G \theta.
\end{split}
\end{equation}
Similarly in the case of interactions with scalar fields, 
spinor fields, etc.
For example, consider a diagram which might appear in resonant pion 
photoproduction at one-loop order shown in figure~\ref{pionfig}.
The parameter dependence of the Green function is contained in the vertices
and the spin-$\frac{3}{2}$ resonance propagator. Our conditions require 
that the vertices and propagators can be written as
\begin{equation}
\begin{split}
\Lambda_{\pi N \Delta}(a) &= \Lambda_{\pi N \Delta} \theta, \\
\Lambda_{\gamma N \Delta}(a) &= \Lambda_{\gamma N \Delta} \theta, \\
\Lambda_{\pi \Delta \Delta}(a) &= \theta \Lambda_{\pi \Delta \Delta} \theta, \\
S_{\Delta}(a) &= \theta^{-1} S_{\Delta} \theta^{-1},\\
S_{\pi}(a) &= S_{\pi},
\end{split}
\end{equation}
where again the $a$ dependence is entirely contained within the $\theta$ factors.
The Green functions are found as in the following example
\begin{equation}
\begin{split}
G(a) &\sim \Lambda_{\pi N \Delta}(a) S_{\Delta}(a) \Lambda_{\pi \Delta \Delta}(a)
S_{\Delta}(a)
S_{\pi}\Lambda_{\pi \Delta \Delta}(a) S_{\Delta}(a) \Lambda_{\gamma N \Delta}(a), \\
&= \Lambda_{\pi N \Delta} \theta \theta^{-1} S_{\Delta} \theta^{-1}  \theta  
\Lambda_{\pi \Delta \Delta} \theta \theta^{-1} S_{\Delta} \theta^{-1} 
S_{\pi}\theta \Lambda_{\pi \Delta \Delta} \theta \theta^{-1} S_{\Delta} \theta^{-1} 
\theta \Lambda_{\gamma N \Delta}, \\
&= \Lambda_{\pi N \Delta} S_{\Delta} \Lambda_{\pi \Delta \Delta} S_{\Delta} S_{\pi}
\Lambda_{\pi \Delta \Delta} S_{\Delta} \Lambda_{\gamma N \Delta}. 
\end{split}
\end{equation}
This example shows the general pattern: all correlation functions for
interacting theories are independent of the arbitrary parameter. 
This is true at every level of perturbation theory and at the 
non-perturbative level as well since the symmetry under $\theta$ is 
non-anomalous as we have discussed above.

\section{Conserved vector current}
\label{implications}

We will now examine one of the ways in which the point transformation invariance 
of the correlation functions can be exploited. 
We will find that the analysis becomes simpler if we reparametrize our general 
action in terms of $b = \frac{a+d}{(1-d)}$ so that the general lagrangian 
(\ref{action1}) becomes \cite{pilling1}
\begin{equation}
\label{Action1}
\mathcal{L} = \ol{\psi}_\alpha 
\left( \Gamma^{\alpha \mu \beta} i \partial_\mu - m \Gamma^{\alpha \beta} \right)
\psi_\beta,
\end{equation}
where 
\begin{equation}
\begin{split}
\Gamma^{\alpha \mu \beta} &= g^{\alpha \beta} \gamma^\mu - A_1 g^{\mu \beta}
\gamma^\alpha - A_2 g^{\mu \alpha} \gamma^\beta + A_3 \gamma^\alpha \gamma^\mu
\gamma^\beta, \\
\Gamma^{\alpha \beta} &= g^{\alpha \beta} - A_4 \gamma^\alpha \gamma^\beta.
\end{split}
\end{equation}
The coefficients are defined in terms of $b$ by
\begin{equation}
\begin{split}
A_1 &= 1 + \frac{(d-2)}{d}b^*, \quad A_2 = 1 + \frac{(d-2)}{d}b \\
A_3 &= 1 + \frac{(d-2)}{d} \left[\frac{(d-1)}{d} |b|^2 + b^* + b \right] \\
A_4 &= 1 + \frac{(d-1)}{d} \left[|b|^2 + b^* + b\right].
\end{split}
\end{equation}
The reason for this reparametrization is that we now have
\begin{equation}
\label{action2}
\mathcal{L} = \ol{\psi}^\alpha \theta_{\alpha \mu}(b^*)
\left(\gamma^{\mu \rho \nu} i \partial_\rho + m \gamma^{\mu \nu}
\right) \theta_{\nu \beta}(b) \psi^\beta \quad
\equiv \ol{\psi}^\alpha \theta_{\alpha \mu}(b^*) \;
\Lambda_{\text{SG}}^{\mu \nu} \; \theta_{\nu \beta}(b) \psi^\beta,
\end{equation}
where the totally antisymmetric combinations of gamma matrices can be 
written as $\gamma^{\mu \rho \nu} = \frac{1}{2} \left( \gamma^\mu \gamma^\rho \gamma^\nu 
- \gamma^\nu \gamma^\rho \gamma^\mu \right)$ and $\gamma^{\mu \nu} = 
\frac{1}{2} \left( \gamma^\mu \gamma^\nu - \gamma^\nu \gamma^\mu \right)$.
The parameter choice $b=0$ now corresponds to the expression, 
$\ol{\psi}^\mu \; \Lambda^{\text{SG}}_{\mu \nu} \; \psi^\nu$,
commonly found \cite{van1981,pascalutsa1999,pascalutsa2003}
as the massive gravitino action in linearized supergravity\footnote{Many 
other expressions are found in the supergravity literature
(for examples see Refs. \cite{madore1975,deser1977,cremmer1978}) but they
are all found to be equivalent to ours by using some choice of parameter $a$ 
and/or using the identity
$\gamma^{\mu \nu \alpha} = i \epsilon^{\mu \nu \alpha \beta} \gamma_5
\gamma_\beta$.}.
The propagator is then
\begin{equation}
\label{propagator2}
S^{\alpha \beta} = \theta^{\alpha \mu}(\tilde{b})
\; S^{\text{SG}}_{\mu \nu} \; \theta^{\nu \beta}(\tilde{b}^*),
\end{equation}
where $S^{\text{SG}}_{\mu \nu}$ is found from (\ref{propagator1}) by
setting $h = 0$. 

Notice that the choice of parameter giving the supergravity action is such 
that the dimension of spacetime, $d$, does not explicitly appear.
This, along with the antisymmetry of $\gamma^{\mu \rho \nu}$ and 
$\gamma^{\mu \nu}$, make many manipulations simpler and many results more 
transparent. Hence, we will use this expression for the general action from now on. 

As we have seen, the observables of the quantum theory are independent of the 
parameter choice. We would like to explore the consequences of this 
invariance. However, the classical action is not invariant under
the transformation since the symmetry transformation $\theta(k)$ of the 
field is in general {\it non-unitary}, $\theta(k^*) \neq \theta^{-1}(k)$.  
Under a global point transformation, 
$\psi_\mu \rightarrow \theta_{\mu \nu}(k)\psi^\nu$, 
the lagrangian (\ref{Action1}) is not invariant, but transforms as 
$\mathcal{L}(b) \rightarrow \mathcal{L}(b \circ k)$. 
In the absence of interactions the equations of motion are invariant since
the free field equations of motion imply that $\gamma \cdot \psi = 0$ and
this makes the point transformation (\ref{pointtrans}) trivial.
However, in the interacting theory this is no longer true. In order to 
explore the consequences of the symmetry we will therefore {\it impose it} 
on the classical action by using the following technique. 
We simply demand that: 
\begin{itemize}
\item{\it classical actions will be considered equivalent if they lead to
the same physical observables}. 
\end{itemize}
Put in another way this says:
\begin{itemize}
\item{\it classical actions will be considered equivalent if they are related 
by a circle-shift redefinition of the parameter}.
\end{itemize}
This makes the point transformations a symmetry of {\it equivalence 
classes} of classical actions and we can examine the consequences. 

Let us re-iterate what we mean by this equivalence to avoid any
possible confusion. We have shown that $\mathcal{L}(b)$ and 
$\mathcal{L}(b \circ k)$ lead to exactly the same physical correlation
functions. These lagrangians are not the same, nor are they
related by any gauge symmetry -- point transformations are {\it not}
gauge transformations. By considering these two different lagrangians 
as being equivalent, we are saying that any interaction whose only 
effect is to change $\mathcal{L}(b)$ into $\mathcal{L}(b \circ k)$,
for some constant $k$, will have no effect on observable physics. 
We now want to examine the consequences of this equivalence.

For simplicity, we will re-write our action so that it is symmetric 
in derivatives\footnote{The reason is so that the analysis results in an 
hermitian current.} and we will restrict the parameter to be real. Thus 
\begin{equation}
\label{symaction}
\mathcal{L}(b) =  \ol{\psi}_\alpha \left[\frac{1}{2} 
\Gamma^{\alpha \rho \beta}(b)i \overset{\leftrightarrow}{\partial}_\rho 
+ m \Gamma^{\alpha \beta}(b) \right] \psi_\beta,
\end{equation}
where we have written 
\begin{equation}
\begin{split}
\Gamma^{\alpha \rho \beta}(b) &=  \theta^{\alpha}_{\; \mu}(b) \gamma^{\mu \rho \nu}
\theta_{\nu}^{\; \beta}(b), \\
\Gamma^{\alpha \beta}(b) &= \theta^{\alpha}_{\; \mu}(b) \gamma^{\mu \nu}
\theta_{\nu}^{\; \beta}(b), \\
\overset{\leftrightarrow}{\partial}_\rho &= 
\overset{\rightarrow}{\partial}_\rho - \overset{\leftarrow}{\partial}_\rho. 
\end{split}
\end{equation}
Under an infinitesimal local point transformation $\theta(k(x))$ the 
lagrangian varies as
$\mathcal{L}(b) \rightarrow \mathcal{L}(b \circ k) + \delta \mathcal{L}$
where $\delta \mathcal{L}$ contains the derivative acting on the parameter
and $\mathcal{L}(b \circ k)$ is defined in exactly the same way as 
$\mathcal{L}(b)$ in (\ref{symaction}) with the derivatives acting only on the fields 
and not on the parameter. Explicit computation gives
\begin{equation}
\begin{split}
\delta \mathcal{L} &= \frac{i}{2d} \ol{\psi}_\alpha \biggl[ \Gamma^{\alpha \rho
\beta} \gamma_\beta \gamma^\nu - \gamma^\alpha \gamma_\beta \Gamma^{\beta
\rho \nu} \biggr] \psi_\nu \left(\partial_\rho k\right).
\end{split}
\end{equation}
Integrating by parts and discarding the surface term we have
\begin{equation}
\label{deltaL}
\mathcal{L}(b) \rightarrow \mathcal{L}(b \circ k) 
- \frac{1}{2d} \left(\partial_\rho J^\rho\right) k(x),
\end{equation}
where $J^\rho$ is given by
\begin{equation}
\label{current}
J^\rho = i \ol{\psi}_\alpha \left[ \Gamma^{\alpha \rho \beta}(b) \gamma_\beta
\gamma^\nu - \gamma^\alpha \gamma_\beta \Gamma^{\beta \rho \nu}(b) \right] \psi_\nu.
\end{equation}
Our symmetry says that $\mathcal{L}(b) = \mathcal{L}(b \circ k)$ in the limit 
that $k(x)$ becomes constant.
This demands that $\delta \mathcal{L} =0$ in the limit of constant $k(x)$. 
Hence from (\ref{deltaL}) we find a conserved current $J^\rho$ associated 
to the global symmetry: $\partial_\rho J^\rho = 0$. 
We see by (\ref{current}) that the current
changes under point transformations by a circle-shift of the parameter
and is therefore invariant according to our symmetry.
We can expand the $\Gamma^{\alpha \rho \beta}$ to find a simpler
expression of the current as follows
\begin{equation}
\begin{split}
J^\rho &= i (1+b) \ol{\psi}_\alpha \left[ \gamma^\alpha g^{\rho \beta} 
- g^{\alpha \rho} \gamma^\beta \right] \psi_\beta, \\
&= i (1+b) \left[\ol{\psi} \cdot \gamma \psi^\rho
- \ol{\psi}^\rho \gamma \cdot  \psi \right].
\end{split}
\end{equation}
Under a transformation $\theta(k)$ the only change is the coefficient 
$(1+b) \rightarrow (1 + b \circ k)$. The conserved charge is given by
\begin{equation}
\label{charge}
Q = i (1+b) \int d^{d-1} x 
\left[ 
\left(\ol{\psi} \cdot \gamma \right) \psi^0 - \ol{\psi}^0 \left(\gamma
\cdot \psi \right) \right].
\end{equation}
The charge can be put in a more suggestive form by defining 
$\chi_1 = \gamma \cdot \psi$ and $\chi_2 = \gamma^0 \psi^0$ leaving
\begin{equation}
\label{newEMcharge}
Q = i (1+b) \int d^{d-1} x 
\left[\chi_1^\dagger \chi_2 - \chi_2^\dagger \chi_1 \right].
\end{equation}
Since we have a conserved current, $J^\rho$, we can couple a 
vector field such as the photon to it as follows
\begin{equation}
\label{newEM}
\begin{split}
\mathcal{L}_{\gamma} &= g J_\mu A^\mu \\
&= i g (1+b) \left[\ol{\psi} \cdot \gamma \psi_\mu
- \ol{\psi}_\mu \gamma \cdot  \psi \right] A^\mu,
\end{split}
\end{equation}
where $g$ is a coupling constant. 
If this coupling is physically reasonable, then it should, among other
things, have a measurable effect on the magnetic moment of the 
spin-$\frac{3}{2}$ particle.
We can also form derivative interactions with scalar fields, such as the 
pion, as 
\begin{equation}
\label{newpion}
\mathcal{L}_{\pi} = g_\pi J^\mu \partial_\mu \phi, 
\end{equation}
where $\phi$ is the scalar field.
Furthermore, we also have the usual conserved vector current coming
from electromagnetic gauge symmetry. This is given by
\begin{equation}
j^\mu = \ol{\psi}_\alpha \Gamma^{\alpha \mu \beta} \psi_\beta,
= \ol{\psi}^\alpha \theta_{\alpha \mu}(b^*)
\gamma^{\mu \rho \nu} \theta_{\nu \beta}(b) \psi^\beta.
\end{equation}
We can use (\ref{Action1}) to write this as
\begin{equation}
\label{oldEM}
\begin{split}
j^\mu &= \ol{\psi}_\beta \gamma^\mu \psi^\beta
- A_1 \left(\ol{\psi} \cdot \gamma \right) \psi^\mu \\
&- A_1 \ol{\psi}^\mu \left(\gamma \cdot \psi \right)
+ A_3 \left(\ol{\psi} \cdot \gamma \right) \gamma^\mu
\left(\gamma \cdot \psi \right),
\end{split}
\end{equation}
where now $A_1 = A_2$ since the parameter is now real.
The definitions of $\chi_1$ and $\chi_2$ allow us to write the
charge as
\begin{equation}
\label{oldEMcharge}
Q_{\text{EM}} = \int d^{d-1} x \left[ \psi_\beta^\dagger \psi^\beta
- A_1 \left( \chi_1^\dagger \chi_2
+ \chi_2^\dagger \chi_1 \right)
+ A_3 \chi_1^\dagger \chi_1 \right],
\end{equation}
and we see that our new symmetry charge (\ref{newEMcharge}) involves only the cross 
terms contained in the usual electromagnetic charge (\ref{oldEMcharge}).

The two currents (\ref{newEM}) and (\ref{oldEM}) are separately 
conserved since they come from independent symmetries and hence we 
can form linear combinations of the two.
Both currents couple to the lower spin components of the vector-spinor 
field and we adjust how much influence these lower spins have. 
It may be possible, with judicious choices of couplings, 
to eliminate the contribution of one or the other of the lower spins 
altogether by eliminating the cross term which contains both 
spin-$\frac{3}{2}$ and spin-$\frac{1}{2}$ components. 
Because of this new freedom, it seems that this new current
will have some influence on the inconsistency problems that have
been found in all interactions involving spin-$\frac{3}{2}$ fields
\cite{johnson1961,deser2000,velo1969}.
Perhaps the inconsistencies can be made to cancel between the
the different conserved currents so that the new symmetry 
can be used to find solutions to that long-standing problem. 
We will turn now to a bit of a review of the consistency 
problem, formulated in $d$-dimensions with arbitrary complex parameter
as well as some ideas about the possibility of solutions (or lack
thereof).

\section{Consistency problems}
\label{consistency}

As we have mentioned in the introductory section, the problem of finding 
consistent interactions for the spin-$\frac{3}{2}$ field is an old one. 
It was first pointed out in paper by Fierz and Pauli in
1939 \cite{fierz1939} where the two-component spinor formalism was used to 
derive lagrangians for massive spin-$\frac{3}{2}$ and spin-2 with
a minimally coupled electromagnetic field. They pointed out that subsidiary
conditions are necessary to reduce the number of independent field components
to the physical number. The subject of spin-$\frac{3}{2}$ was revisited by 
Rarita and Schwinger in 1941 \cite{rarita1941} who developed the notation and 
action that are now most often used. They also noticed that the theory 
was not unique and that a collection of actions all give equivalent theories. 
This is the same non-uniqueness that we have exploited in the present paper.

Consistency problems with the theory were further discussed in 1961 by Johnson 
and Sudarshan (JS) \cite{johnson1961} who showed that in the presence of an 
electromagnetic field the field anti-commutator becomes indefinite, 
i.e. a Lorentz frame can always be found in which it is negative. 
A related discovery by Velo and Zwanziger (VZ) in 1969 \cite{velo1969} was 
that there are modes of the field which propagate faster than light. 
These problems derive from precisely the reasons given by Fierz and Pauli in 1939, 
namely that the lower spin particles should be removed by subsidiary conditions or 
they will give rise to negative energy states and indefinite
charges\footnote{However, see \cite{kirchbach2001} for an interesting
alternative to this.} and the number of subsidiary conditions must remain 
invariant to the presence of interaction. 

In higher spin field theories involving auxiliary components and
constraints, the time derivative operator of the lagrangian is a singular matrix. 
That this is necessary can be exemplified with our spin-$\frac{3}{2}$ operator in 
(\ref{Action1}). The dynamical term, which contains the time derivative operator 
in the equations of motion, is given by
\begin{equation}
\label{singop}
\Gamma^{\alpha 0 \beta} i \partial_0 \psi_\beta,
\end{equation}
where $\Gamma^{\alpha 0 \beta}$ has suppressed spin indices and can be
viewed, in 4 dimensions, as a $16 \times 16$ matrix. 
The equations of motion imply a set of constraints (equations among the field 
components which do not contain time derivatives) and so not all of the components 
of $\psi_\beta$ are dynamical and these non-dynamical components will
not appear in (\ref{singop}). Hence, there must exist vectors, corresponding to 
the non-dynamical components, which are annihilated by the above matrix.
Therefore the matrix is necessarily singular and will have a determinant 
of zero. 

In the case of spin-$\frac{3}{2}$, the constraints implied by the equations
of motion are: the primary constraint, found by applying 
$\theta^{0}_{\; \alpha}(\tilde{b^*})$ to the equations of motion (\ref{eqmot}), 
and the secondary constraint, found by taking the covariant derivative of 
the equations of motion. 
In four spacetime dimensions, as we have mentioned, these constitute 
a total of eight constraints needed to reduce the 16 component 
vector-spinor field to $2(2s+1)=8$ onshell degrees of freedom. 
In order to obtain a non-singular operator in (\ref{singop}) we must impose
the constraint equations and eliminate the non-dynamical components in
terms of dynamical ones, thus giving the on-shell equations of motions in 
terms of a new non-singular matrix. 

In the free field case everything works perfectly and the non-dynamical 
components can be eliminated without trouble. However, a problem occurs with 
the constraints in the presence of interaction in that the secondary constraint 
becomes dependent on the external field. This fact can lead to inconsistencies
as we will now show for the canonical example of electromagnetic minimal coupling.
Our general action in the case of electromagnetic minimal coupling is
found by inserting a covariant derivative into (\ref{action2}), giving
the following equations of motion
\begin{equation}
\label{eqmot}
\theta_{\alpha \mu}(b) \left(\gamma^{\mu \rho \nu} i D_\rho + m \gamma^{\mu \nu}
\right) \theta_{\nu \beta}(b) \psi^\beta = 0,
\end{equation}
where $D_\rho = \partial_\rho + i e A_\rho$ is the covariant derivative.
If we apply $\theta^{0 \alpha}(\tilde{b})$ to (\ref{eqmot}) we are left with an 
expression which contains no time derivatives (since $\gamma^{\mu \rho \nu}$ vanishes 
when two indices are zero) and is the primary constraint.
To derive the secondary constraint we apply $\gamma^\alpha$ to (\ref{eqmot})
to get the `useful relation'
\begin{equation}
\label{rel}
i D \cdot \psi = \left( i \frac{(d-1)b + d}{d} \Nt{D} + \frac{(d-1)(1+b)}{d-2} m
\right) \gamma \cdot \psi.
\end{equation}
Now take the covariant derivative of (\ref{eqmot}), using (\ref{rel}) 
as well as the identities
\begin{equation}
\label{ident}
\begin{split}
\gamma_{\mu \nu} D^\mu D^\nu &= \frac{ie}{2} \gamma_{\mu \nu} F^{\mu \nu}, \\
\Nt{D} \Nt{D} &= \frac{ie}{2} \gamma_{\mu \nu} F^{\mu \nu} + \frac{1}{2} D^2, 
\end{split}
\end{equation}
to arrive at the secondary constraint
\begin{equation}
\label{const2}
\left( \frac{ie}{2} \frac{(d-2)b + d}{d} \gamma_{\mu \nu} F^{\mu \nu} 
+ \frac{(d-1)(1+b)}{d-2} m^2 \right) \gamma \cdot \psi 
= ie \gamma_\mu F^{\mu \nu} \psi_\nu.
\end{equation}
In four\footnote{A similar identity will hold in any 
even dimension where there exists a $\gamma^5$--like matrix \cite{wetterich1983}.}
spacetime dimensions we can use the 
relation\footnote{We are using a mostly minus metric, diagonal $\gamma^0$,
and $\gamma^5 = -i \gamma^0 \gamma^1 \gamma^2 \gamma^3$.} 
\begin{equation}
\label{ident2}
\frac{1}{2} \gamma_{\mu \nu} F^{\mu \nu} \gamma \cdot \psi
= \gamma_\mu F^{\mu \nu} \psi_\nu - i \gamma^5 \gamma_\mu \star F^{\mu \nu} \psi_\nu,
\end{equation}
with the dual field strength
$\star F^{\mu \nu} 
= \frac{1}{2} \epsilon^{\alpha \beta \mu \nu} F_{\alpha \beta}$
to write (\ref{const2}) in the simpler form
\begin{equation}
\label{const2b}
\gamma \cdot \psi 
= - \frac{ie}{3(1+b)m^2} \gamma_\mu \mathcal{F}^{\mu \nu} \psi_\nu,
\end{equation}
where 
\begin{equation}
\label{F}
\mathcal{F}^{\mu \nu} = b F^{\mu \nu} + i (b + 2) \gamma^5 \star F^{\mu \nu}.
\end{equation}
It should now be clear that the free field constraint $\gamma \cdot \psi = 0$ is no 
longer true unless the interaction vanishes. 

Now that we have the constraints in the presence of electromagnetic
minimal coupling we can describe the general case of the well known 
problem \cite{deser2000}. 
The secondary constraint (\ref{const2b}) can be written as
\begin{equation}
\label{const2c}
\left(\gamma^0 +  \frac{ie}{3(1+b)m^2} \gamma_\mu \mathcal{F}^{\mu 0} \right) \psi^0 
= \left( \gamma^k + \frac{ie}{3(1+b)m^2} \gamma_\mu \mathcal{F}^{\mu k} \right) \psi^k,
\end{equation}
and one can easily see from the equations of motion (\ref{eqmot}) that, in the case
$b=0$, the components $\psi^0$ are non-dynamical. In that case they must be 
fixed by the above condition so that they can be eliminated in terms of the 
dynamical field components. 
To do this, one must solve (\ref{const2c}) for $\psi^0$. This requires that the matrix
$\left(\gamma^0 +  \frac{ie}{3(1+b)m^2} \gamma_\mu \mathcal{F}^{\mu 0} \right)$
be non-singular. This implies, since $b=0$, that the matrix 
$\left( 1 - \frac{2e}{3m^2}\gamma^0 \vec{\gamma} \cdot \vec{B} \gamma^5 \right)$
is non-singular, which in turn requires that
$\left( 1 + \frac{2e}{3m^2} \vec{\sigma} \cdot \vec{B} \right)$
is non-singular.
Taking the determinant shows that this latter matrix is singular when 
$|\vec{B}|^2 = \left(\frac{3m^2}{2e}\right)^2$. This is the famous 
result of \cite{johnson1961} and \cite{velo1969} and 
means that if this equation is used to eliminate $\psi^0$ in the
classical equations of motion one would find that, at this value of the
external field, not all of the components of $\psi^0$ are determined and
these undetermined components lead to space-like characteristic surfaces
and the possibility of field modes propagating acausally.
These non-physical characteristic surfaces indicate that the number 
of constraints has changed and that non-dynamical field components are 
still present in the field equations. For example, the operator
coefficient to the time derivative in the equations of motion will still
be a singular matrix at this external field value.

Note that we have conducted the analysis for $b = 0$, but we have lost
no generality since the point transformation invariance can be used to 
transform this to the case of arbitrary $b$. 
To see this, Let us briefly outline the method \cite{velo1969,madore1973} that is 
normally used to analyze the consistency of the classical field equations. 
The basic idea is to look at the characteristic 
surfaces given by the classical on-shell differential equation. To do this one
plugs the constraints into the equations of motion, thus (hopefully) eliminating 
all of the non-dynamical degrees of freedom. Then the characteristics are found 
by replacing the derivative in the equation with a four-vector $n^\mu$. This
vector will be a normal to a characteristic surface (a surface along which the 
maximum velocity solutions to the differential equation are restricted to propagate) 
if it is the solution to a certain equation (\ref{characteristics}). If we want only
time-like solutions, so that causality is preserved, we would like the normals
to the characteristic surfaces to remain always space-like. 

The normals to the characteristic surfaces $n_\mu$ are given by the determinant
\begin{equation}
\label{characteristics}
D(n) = \left| \tilde{\Gamma}^{\alpha \mu \beta} n_\mu \right| = 0,
\end{equation}
where we have replaced the derivatives in the field equations with
$n_\mu$ and $\tilde{\Gamma}^{\alpha \mu \beta}$ is the operator in the field 
equations {\it after all of the constraints have been imposed}.
If there is a time-like normal we can use Lorentz invariance to write 
it as $n_\mu = (n,0,0,0)$. Our characteristic equation is then
\begin{equation}
\label{det}
D(n) = n^{16} \left| \tilde{\Gamma}^{\alpha 0 \beta} \right| = 0.
\end{equation}
Notice that if this determinant vanishes for $b=0$, then it also vanishes
for arbitrary $b$ by the properties of the point transformations.
The vanishing of the determinant means that there are characteristic surfaces which are 
space-like, indicating the possibility that field components propagate acausally. 
We say `indicating {\it the possibility}' of acausal propagation because if
the determinant was zero, so that there are space-like characteristics, one would still 
have to prove that there were actual physical field modes that propagate along these 
acausal characteristics. This is usually done with the method of shock
discontinuities \cite{madore1973}. We will not discuss this method further since it 
has been established many times that there are field modes which propagate acausally
in various cases. On the other hand, to prove that a theory is consistent one would 
only need to show that there are no space-like characteristic surfaces and it
would follow that there is no acausal propagation, without need of the method 
of discontinuities.

Notice that if we didn't impose constraints, the matrix in the determinant 
equation (\ref{det}) would be the same one that we discussed earlier
(\ref{singop}) which is necessarily singular.
Thus, if the constraints failed to eliminate a non-dynamical field component. This
component would remain in the field equations but it would, by definition, 
not appear in the time derivative part since it is non-dynamical.
Hence the matrix coefficient of the time derivative $\partial_0$ would annihilate 
the 16-component vector representing this component. The matrix then
annihilates a non-zero vector and so it must have determinant equal to zero.
Conversely, the physical components all have time derivatives which means
the matrix coefficient of the time derivative does not annihilate them. 

It is interesting that the inconsistency comes from the breakdown of
constraints and is caused by non-dynamical, `unphysical' components creeping back 
into the field equations at certain values of the external field and the presence
of these components leads to physical information propagating
acausally. Therefore, if we could find a covariant method of preventing
these extra components from coming back the problems would be solved. 
Unfortunately, this seemingly 
simple task is a subtle one and has not been accomplished in general since the problem 
was discovered over 40 years ago although there have been many papers 
written on the subject\footnote{We should mention the recent 
developments on a new type of spin 1/2 field with mass dimension one 
(called `Elko') given in reference \cite{ahluwalia2004}. That work 
can be extended to spin 3/2 and may lead to new results for the 
consistency problem.}
\cite{hagen1971,singh1973,hortacsu1974,prabhakaran1975,capri1980,aurilia1980,sierra1982,darkhosh1985}.
We do not claim to have solved the problem here either, but we will present an idea that
may be of some use in the search for a solution or a proof that one can't
exist.

We should mention that, in the context of supergravity, when the mass of the 
spin-$\frac{3}{2}$ field is tuned in certain ways to the background spacetime, 
there are theories which seem to be free of inconsistencies \cite{deser2001}.
On the other hand, if we would like to use the theory to effectively model low mass 
spin-$\frac{3}{2}$ fields (such as the $\Delta(1232)$ nucleon resonance) at 
low energies one cannot expect the background spacetime or the 
existence of supersymmetry partners to be of much help. 

We have exemplified the problem for the case $b=0$ (i.e. the supergravity
action) but, as we have said, the same is true for arbitrary $b$ since the 
lagrangian operators are related by a non-singular point transformation
and so if the determinant (\ref{characteristics}) is zero for one value of the 
parameter, a point transformation will not alter this and it will be zero for 
any parameter value.

In any case, we can now ask: is it possible that the new symmetry current 
and charge can help us? The conserved charge arises from a symmetry of the path 
integral which is not shared by the usual classical theory, so if a solution 
to this problem could be found from this charge it would have the happy consequence 
that the inconsistency which is present in the classical theory would not affect 
the physical correlation functions. Conversely, it is also possible that the extra 
symmetry of the path integral may make things even worse.

Suppose we begin with no interaction and then slowly turn
on an external field. In that case we can argue that the charge should 
remain zero since it is zero in the free case and is conserved. 
The charge (\ref{charge}) is given in an electromagnetic background by
\begin{equation}
\label{interactingcharge}
Q = i (1+b) \int d^3 x \left[ 
\ol{\psi}_\nu \left(\mathcal{F}^{\nu \mu}\right)^* \gamma_\mu \psi^0 
- \ol{\psi}^0 \gamma_\mu \mathcal{F}^{\mu \nu} \psi_\nu
\right]. 
\end{equation}
The conservation of charge is related to the conservation of constraints
since the spin-$\frac{1}{2}$ fields $\gamma \cdot \psi$ and $\psi^0$
in the charge are exactly the ones which are usually removed via the
constraint structure of the free field Rarita-Schwinger equations. Hence it seems
possible that the loss of constraints, which signals the onset of the
inconsistency, will have a direct effect on the charge. This gives us an
idea for a possible new direction. When the external field is such that there 
is a loss of constraints the charge should remain unaffected since it is conserved. 
It is possible then, that in that case, the equation for the charge would represent 
an additional constraint which could be used to eliminate the unphysical
fields which appear when the usual constraints breakdown. 
However, we do not want the extra constraint in general, but only in the case 
when the usual constraints breakdown and then only so as to exactly 
compensate for the breakdown. This seems like quite a demand. 
If the conservation of charge implied additional contraints in
general we would then have too many constraints as soon as the background 
is turned on. So physically we want the charge to be identically 
zero when the constraints are imposed. The only time the situation 
changes is when the usual constraints break down. In that case we want the 
charge to no longer vanish identically, but to itself become a constraint
compensating for the ones that were lost. The usual constraints are given by
$\gamma \cdot \psi \sim \gamma_\mu \mathcal{F}^{\mu \nu} \psi_\nu$ and this
combination then also appears in the expression for the charge. So
perhaps it is possible that a breakdown of this equation would have a compensating 
effect in the conserved charge. 
In any case, we will leave this problem for now with the hope that these
thoughts may inspire new angles of attack so that either a new solution, or
a new reason for the lack of one, can be found.

\section{Conclusion}

We will now summarize the main features of our work.
We have derived the general lagrangian and propagator for the 
Rarita-Schwinger field in $d$-dimensions. These are given by equations 
(\ref{Action1}) and (\ref{propagator1}) respectively and should prove 
useful in calculating higher loop effects in dimensional regularization
as would occur in the effective resonance contribution to the imaginary 
part of pion scattering amplitudes, anomalous magnetic moments, and 
many other processes for which the $\Delta(1232)$ resonance or any other 
spin-$\frac{3}{2}$ particles play a significant role.

We studied the point transformation algebra and explored the invariance 
properties of the general action under rotations of the lower spin, 
off-shell fields. We found that this invariance implies the existence 
of a conserved vector current and charge. The conserved current leads 
to interactions involving spin-$\frac{3}{2}$ fields such as the 
electromagnetic couplings that we have given in (\ref{newEM}) above, 
possible couplings to other vector fields such as vector mesons, 
derivative couplings to scalar fields such as the pion, etc. 
It is important to check the predictions of these interactions. 

Finally, we looked at the consistency problems and indicated two possible
avenues where progress might be made. The first is a possible
cancellation of the problematic terms by tuning interactions based
on the conserved current. The second is by using the conserved
charge as an additional constraint to compensate for the loss of the usual
constraints at the `bad' values of the external magnetic field.

It is important to attempt a generalization of these techniques in the 
theory of higher spin fields in the same way as done for spin-$\frac{3}{2}$. 
It seems likely that similar (though more complicated) groups will exist 
which rotate among the auxiliary lower spin components in those cases as well, 
leading again to new conserved currents. 

\section*{Acknowledgments}

I would like to thank Emil Akhmedov and D. V. Ahluwalia-Khalilova 
for discussion and helpful comments.

\end{document}